\documentclass[lettersize,journal]{IEEEtran}

\usepackage{formatting/packages}
\usepackage{formatting/macros}

\begin{document}
\captionsetup{skip=0pt}
\setlength{\abovecaptionskip}{1pt}
\setcounter{topnumber}{5}

\title{\LARGE{Understanding the Performance Behaviors of \\End-to-End Protein Design Pipelines on GPUs}}

\author{Jinwoo Hwang, Yeongmin Hwang, Tadiwos Meaza, Hyeonbin Bae, \IEEEmembership{Member, IEEE}\\
Jongse Park, \IEEEmembership{Senior Member, IEEE}

\thanks{Manuscript received 13 November 2025; accepted 16 December 2025. Date of publication 19 December 2025; date of current version 6 January 2026. Digital Object Identifier 10.1109/LCA.2025.3646250. This work was supported by the National Research Foundation of Korea (NRF) grant funded by the Korea government (MSIT) (RS-2024-00342148).}
\thanks{The authors are with the School of Computing, Korea Advanced Institute of Science and Technology, Daejeon, 34141, South Korea. (e-mail: {jwhwang,ymhwang,tadiwos,hbbae,jspark}@casys.kaist.ac.kr).}
\vspace{-3ex}
}

\markboth{IEEE Computer Architecture Letters,~Vol.~25, No.~1, January~2026}
{Hwang \MakeLowercase{\textit{et al.}}: Understanding the Performance Behaviors of End-to-End Protein Design Pipelines on GPUs}

\maketitle
\begin{abstract}
Recent computational advances enable protein design pipelines to run end-to-end on GPUs, yet their heterogeneous computational behaviors remain undercharacterized at the system level.
We implement and profile a representative pipeline at both component and full-pipeline granularities across varying inputs and hyperparameters.
Our characterization identifies generally low GPU utilization and high sensitivity to sequence length and sampling strategies.
We outline future research directions based on these insights and release an open-source pipeline and profiling scripts to facilitate further studies.
\end{abstract}


\IEEEpubidadjcol
\vspace{-0.1cm}
\section{Introduction}
\label{sec:intro}
\IEEEPARstart{P}{roteins} are the fundamental molecular machinery driving biological systems.
Recent advances in protein AI models~\cite{alphafold3, rfdiffusion, proteinmpnn, esm-2, mmseqsgpu, protenix, diffdock, nature-colabfold:2022} have significantly reduced the cost of computational protein structure prediction, enabling the design of proteins with enhanced or novel functionalities relevant to next-generation therapeutics, industrial biocatalysts, and biosensors~\cite{nature-opportunities:2024}.
Protein design pipelines typically involve multiple stages with distinct computational behaviors and heavy GPU dependencies, yet their combined system-level performance implications remain largely unexplored.

We address this gap by systematically profiling a representative GPU-based protein design pipeline at both component and pipeline granularities, examining performance sensitivity to sequence length, hyperparameters, and sampling strategies.
Our study investigates GPU utilization patterns and the impacts of GPU scheduling under co-location scenarios.
We also outline directions for future research, including developing cost-aware metrics, scaling to multi-GPU environments, and exploring agent-assisted orchestration.
To foster reproducibility and accelerate future studies, we release our pipeline and profiling tools as open-source software.~\footnote{\url{https://github.com/casys-kaist/protein-design-pipelines.git}}

\vspace{-0.2cm}
\IEEEpubidadjcol
\section{Background and Motivation}
\subsection{Proteins and In Silico Screening}
\niparagraph{Protein engineering.}
The diversity of protein function arises from their three-dimensional structures encoded by amino acid sequences.
Changing the order and composition of amino acids can therefore alter a protein's function.
Historically, discovery and optimization heavily relied on physical screening and structural characterization, which are costly and slow, limiting exploration of the protein design space.

\niparagraph{AI-driven in silico screening.}
Recent advancements in AI have transformed this workflow by providing reliable structural and functional predictions at scale.
With these models, the marginal cost of evaluating designs computationally is significantly lower compared to traditional wet-lab experiments.
Consequently, design cycles that previously took months now require only hours using modest GPU clusters, paving the way for a new paradigm of rapid, in silico screening.

\niparagraph{High-throughput screening.}
This paradigm involves rapidly evaluating extensive virtual libraries of protein candidates.
Early computational filters assess candidates based on functionality, manufacturability, and stability, eliminating unsuitable designs early.
The remaining high-priority candidates receive focused experimental attention, advancing only the strongest designs to physical validation.
Under such conditions, high-throughput in silico screening becomes critical.
Large batches increase the exploration of the design space and enhance the probability of identifying superior candidates~\cite{massivefold}.

\vspace{-0.3cm}
\subsection{Workflow of Protein Design}
\label{subsec:protein-design-pipelines}
\begin{figure}[t]
  \centering
  \includegraphics[width=\linewidth]{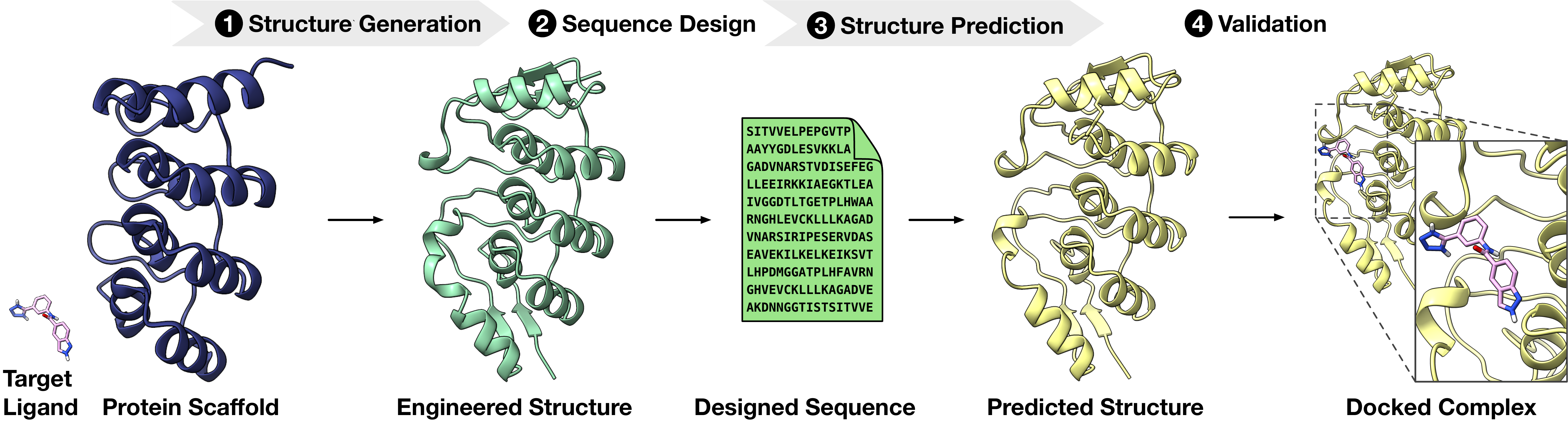}
  \caption{Four major stages of a protein-design pipeline.}
  \vspace{-3ex}
  \label{fig:overview}
\end{figure}
Protein design workflows vary depending on the specific objectives and are assembled from interchangeable components.
Nonetheless, they commonly follow a four-stage pattern, as depicted in Figure~\ref{fig:overview}.

\circledparagraph{1}{Structure generation.}
Structure generation proposes engineered protein structures satisfying geometric or functional constraints, using models such as RFdiffusion~\cite{rfdiffusion}.
Structures can be designed from scratch or derived from a known scaffold protein, often specifying target ligands for optimization.

\circledparagraph{2}{Sequence design.}
Sequence design generates amino acid sequences that match the given structures, using inverse folding models such as ProteinMPNN~\cite{proteinmpnn}.
Protein language models, including ESM-2~\cite{esm-2}, further refine and explore sequences.

\circledparagraph{3}{Structure prediction.}
Structure prediction folds sequences back into three-dimensional structures with predictors like AlphaFold 3~\cite{alphafold3}, ESMFold~\cite{esm-2}, and Protenix~\cite{protenix}.
Most models other than ESMFold require evolutionary context, performing homology searches and sequence alignments via MMseqs2~\cite{mmseqsgpu}.

\circledparagraph{4}{Validation.}
Evaluation estimates key properties, such as binding affinity and structural stability, by forming docked complexes and applying scoring methods like AutoDock Vina and DiffDock~\cite{vina,vina-gpu,diffdock}.
For comprehensive context and tool comparisons, refer to a recent survey~\cite{nature-primer:2025}.

\subsection{Heterogeneity across Components}
\niparagraph{Diverse workloads.}
Components employ varied model architectures and kernels, ranging from Diffusion~\cite{rfdiffusion, diffdock, alphafold3, protenix}, Transformers~\cite{esm-2,alphafold3,protenix}, graph neural networks~\cite{proteinmpnn}, custom GPU kernels for sequence alignment~\cite{mmseqsgpu}, and scoring methodologies~\cite{vina-gpu}.
This architectural diversity results in distinct computational patterns, activation footprints, and kernel behaviors.
Inputs and hyperparameters further amplify this diversity.
Some components are sensitive to sequence length, and several expose adjustable hyperparameters balancing throughput and accuracy, such as diffusion step counts.

\niparagraph{Need for characterization.}
These insights motivate two complementary analyses.
Initially, we characterize individual components across various inputs and hyperparameters.
Subsequently, we evaluate the complete end-to-end pipeline performance on shared GPUs, analyzing scheduling interactions and performance implications at the system level.

\vspace{-0.1cm}
\section{Protein Design Pipeline}
\subsection{Protein Engineering Scenario}
We illustrate a protein engineering scenario wherein a stable, well-characterized scaffold protein is modified to engage a specific small-molecule target ligand.

\niparagraph{Protein and ligand pairs.}
\newcommand{\idsize}[3]{\makecell[c]{#1\\ #2 \textsc{#3}}}
\begin{table}[t]
\centering
\caption{Selected scaffolds and ligands. We report PDB IDs with lengths: amino acid residues (aa) and heavy atoms (ha).}
\label{tab:dataset}
\begin{tabular}{cccc}
\toprule
 & Short & Medium & Long \\
\midrule
Scaffold ID (\textsc{aa}) & 1SHG (62) & 1N0R (126) & 1TIM (247) \\
Ligand ID (\textsc{ha}) & 3DX1 (9)  & 4DE1 (23)  & 3PRS (50) \\
\bottomrule
\end{tabular}
\vspace{-3ex}
\end{table}

We select four protein scaffolds and three ligands to represent a diverse range of sequence lengths and complexities, as detailed in Table~\ref{tab:dataset}.
For protein scaffolds, we use well-established structures spanning short to long sequences.
For ligands, we sample three molecules from the CASF-2016~\cite{casf-2016} dataset, choosing examples at the 1st, 50th, and 99th percentiles by length.

\niparagraph{Components and hyperparameters.}
We select widely adopted models for each pipeline component to reflect current industry standards, as outlined in Section~\ref{subsec:protein-design-pipelines}.

\begin{table}[t]
\centering
\caption{Components and hyperparameters. Thr./Bal./Acc. denote throughput-optimized, balanced, and accuracy-oriented.}
\label{tab:knobs}
\small
\setlength{\tabcolsep}{4pt}
\begin{tabular}{lllll}
\toprule
Component & Hyperparameter & Thr. & Bal. & Acc. \\
\midrule
RFdiffusion & \texttt{num\_steps} & 20 & 50 & 100 \\
ProteinMPNN & -- & -- & -- & -- \\
ESM-2 & \texttt{model\_size} & 150M & 650M & 3B \\
MMseqs2 & \texttt{max\_iterations} & -- & 3 & 5 \\
Protenix & \texttt{num\_steps} & 100 & 200 & 300 \\
Protenix & \texttt{num\_cycles} & 2 & 4 & 8 \\
ESMFold & \texttt{num\_cycles} & 2 & 4 & 8 \\
AlphaFold3 & \texttt{num\_steps} & 100 & 200 & 300 \\
AlphaFold3 & \texttt{num\_cycles} & 5 & 10 & 15 \\
DiffDock & \texttt{num\_steps} & 10 & 20 & 30 \\
Vina-CPU & \texttt{exhaustiveness} & 4 & 8 & 16 \\
Vina-GPU & -- & -- & -- & -- \\
\bottomrule
\end{tabular}
\vspace{-3ex}
\end{table}

Regarding hyperparameters, we establish three configurations: \emph{throughput-optimized}, \emph{balanced}, and \emph{accuracy-oriented}.
The balanced configuration aligns with default settings from reference implementations whenever possible, while the other configurations adjust computational resources to balance throughput and quality.
Table~\ref{tab:knobs} summarizes these hyperparameters for each component.

\niparagraph{Scope and limitations.} This synthetic scenario aims to characterize system performance comprehensively. It captures size-driven impacts and heterogeneity across pipeline components, spanning representative computational conditions under controlled variations in protein and ligand sizes. However, it does not evaluate biochemical compatibility or the experimental viability of specific scaffold-ligand pairs.

\subsection{Implementation Details}
\label{subsec:implementation-details}
\niparagraph{Software setup.}
Each component is containerized with Docker to ensure reproducibility and portability.
Pipeline workflows are orchestrated using Nextflow~\cite{nextflow}, a widely adopted bioinformatics workflow manager.
Tasks are scheduled through Kubernetes, augmented by the Alibaba GPUShare plugin, which exposes GPU memory as a schedulable resource to prevent out-of-memory errors during concurrent operations.
Empirically, we observe negligible performance impact from these orchestration layers given our workload characteristics.
The software stack comprises Ubuntu 22.04, Nextflow 25.04, Docker 28.1, and Kubernetes 1.33.

\niparagraph{Hardware setup.}
Experiments are conducted on a server equipped with one RTX~3090 GPU, two Intel Xeon Gold~6326 CPUs, 256~GB DRAM, and 4~TB PCIe Gen~4 NVMe SSD.

\vspace{-0.1cm}
\section{Performance Characterization}
\subsection{Component-Level Observations}
\label{subsec:component-level-observations}
\begin{figure*}[t]
\centering
\includegraphics[width=\textwidth]{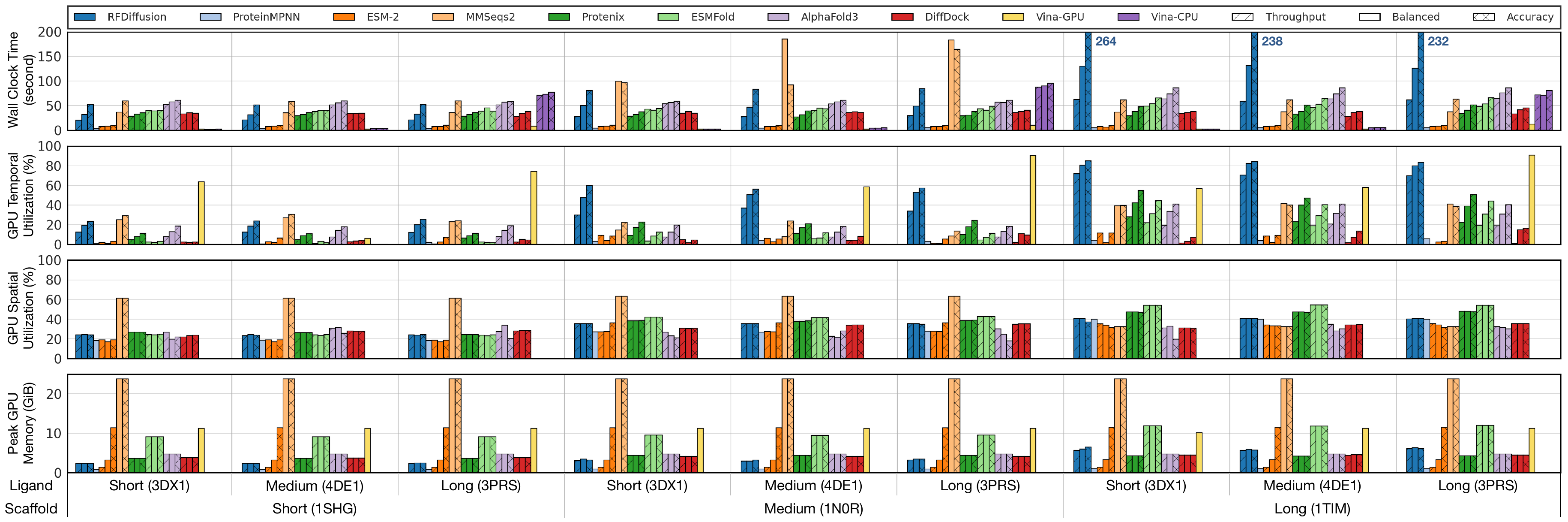}
\caption{Profiling of individual pipeline components across varying inputs and hyperparameters. Vina-CPU does not utilize GPU resources. Vina-GPU is implemented using OpenCL, which does not support spatial profiling; thus, only temporal utilization is reported.}
\vspace{-3ex}
\label{fig:microbenchmark}
\end{figure*}
Figure~\ref{fig:microbenchmark} profiles individual components of the protein-design pipeline across three operating points and nine scaffold–ligand pairs spanning different sequence lengths.
We measure wall-clock time, temporal utilization, and peak device memory usage via the NVIDIA Management Library (NVML).
To evaluate spatial utilization, we first aggregate kernels accounting for at least 90\% of the execution time using Nsight Systems, then calculate the runtime-weighted average SM occupancy with Nsight Compute.

\niparagraph{Inputs and hyperparameters.}
Input length significantly influences runtime across components.
RFdiffusion exhibits a particularly sharp increase, with runtime scaling up to 5.1~$\times$ when scaffold length expands from 1SHG to 1TIM. Other components experience noticeable but smaller runtime increases.
MMseqs2 shows variable runtime spikes driven by factors beyond sequence length alone, including database interactions.
Hyperparameter adjustments consistently influence runtime, although input length generally has a more substantial effect.
Accuracy-oriented configurations extend execution time compared to throughput-oriented settings.

\niparagraph{GPU utilization.}
Temporal GPU utilization varies significantly across components.
RFdiffusion and Vina-GPU achieve relatively high utilization.
MMseqs2 shows moderate utilization around 26\%. Vina-GPU utilization notably improves for larger ligands, reaching 70 to 91\%.
RFdiffusion similarly achieves higher utilization with longer sequences.
ESMFold, Protenix, and AlphaFold~3 occasionally reach moderate utilization with larger inputs, yet low utilization dominates overall.

Spatial utilization is generally modest, predominantly clustering between 24\% and 38\%.
Notable exceptions include ESMFold’s large input scenario, which reaches approximately 54\%, and MMseqs2, consistently higher at 61 to 63\%.
Experiments conducted on an RTX 3090 suggest the model sizes and execution parameters do not fully leverage the GPU’s computational capacity, indicating room for optimization.

\niparagraph{Peak GPU memory.}
Peak GPU memory usage remains consistently below device limits of 24~GB, except for MMseqs2, suggesting ample headroom for co-locating multiple processes.

\subsection{Pipeline-Level Observations}
\label{subsec:pipeline-level-observations}
We extend our analysis from isolated components to entire end-to-end pipelines, using the balanced configuration with medium-length scaffolds and ligands.
For the structure prediction stage, we select Protenix, and for the validation stage, we use Vina-GPU.

\niparagraph{Sampling for exploration.}
Several components offer sampling hyperparameters that expand the search space.
RFdiffusion can produce multiple backbone structures per prompt, ProteinMPNN can generate multiple sequences per backbone, ESM models propose sequence variants, AlphaFold-style predictors return multiple structural candidates, and AutoDock evaluates several docking poses.
When an upstream component outputs multiple candidates, these propagate downstream.

\begin{figure}[t]
\centering
\vspace{-1ex}
\includegraphics[width=\linewidth]{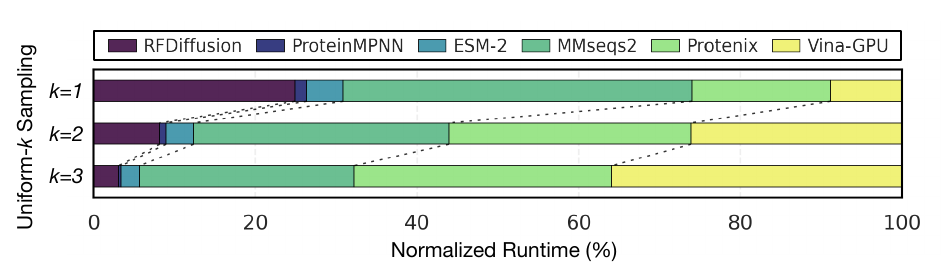}
\vspace{-4.5ex}
\caption{Normalized runtime of pipeline components across varied sampling.}
\vspace{-3ex}
\label{fig:fanout}
\end{figure}
To illustrate the impact of sampling, we experiment with a uniform sampling configuration, setting sampling counts to 1, 2, or 3 for all components supporting sampling, producing 1, 16, and 81 samples, respectively.
Figure~\ref{fig:fanout} shows the normalized runtime breakdown for each case.
With a single sample per component, the total runtime aligns with the sum of component-level runtimes shown in Figure~\ref{fig:microbenchmark}.
Reusing common outputs from early stages significantly improves exploration throughput, reducing per-sample latency from 4.32 minutes (uniform-1) to 1.18 minutes (uniform-2) and further down to 0.83 minutes (uniform-3).
As sample counts increase, runtime shifts towards downstream components, reflecting multiplicative workload expansion.

\niparagraph{GPU co-location.}
\begin{figure*}[t]
\centering
\includegraphics[width=\textwidth]{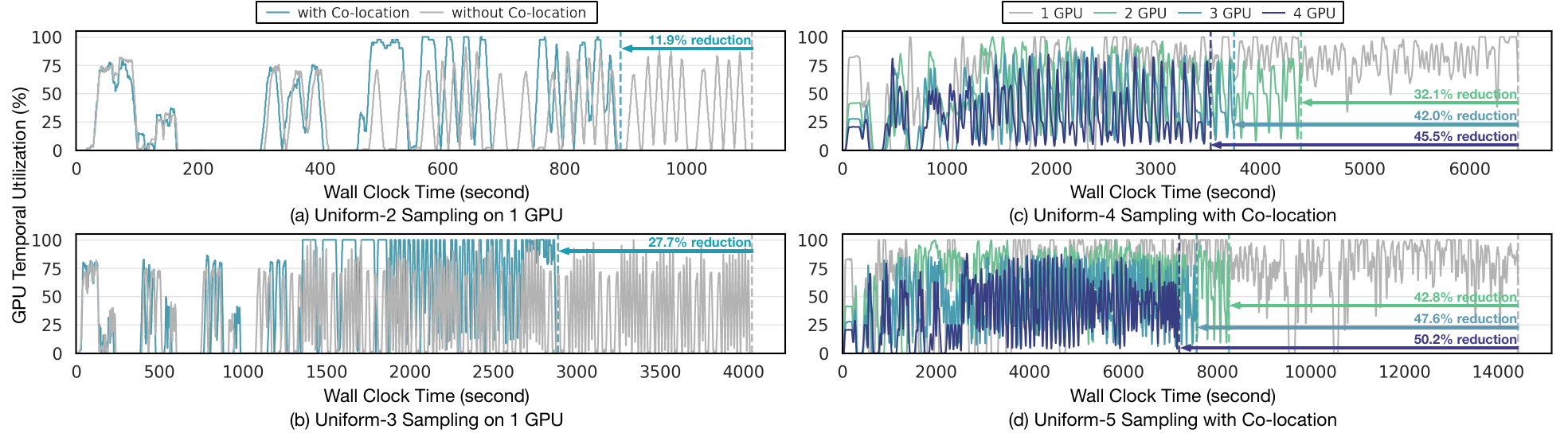}
\caption{GPU temporal utilization over execution time.}
\vspace{-3ex}
\label{fig:pipeline-level}
\end{figure*}
Figure~\ref{fig:pipeline-level}(a) and (b) illustrate GPU temporal utilization with and without co-location on a single GPU, for uniform sampling counts of two and three, respectively.
Co-location, in which multiple processes timeshare GPU resources, enhances temporal utilization and throughput, reducing overall execution time.
This benefit grows with higher sampling counts, providing greater scheduling flexibility.
Specifically, co-location decreases latency by 11.9\% for uniform-2 sampling and by 27.7\% for uniform-3 sampling.
Later-stage components particularly benefit from co-location due to their lower peak memory requirements, allowing more concurrent executions.

\niparagraph{Multi-GPU scaling.}
Figures~\ref{fig:pipeline-level}(c) and (d) evaluate scaling to multiple GPUs using higher uniform sampling counts of four and five with co-location enabled.
In early stages, limited sample availability causes low mean temporal utilization as many GPUs remain idle.
Utilization gradually improves over time as more samples become available.
At current uniform sampling levels, GPUs remain generally underutilized, resulting in weak scaling efficiency.
Higher sampling counts lead to stronger scaling improvements.
For example, scaling from one to two GPUs reduces latency by up to 42.8\% with uniform-5 sampling but shows less improvement with uniform-4 sampling.
Scaling efficiency noticeably declines at three and four GPUs with latency reductions reaching only 47.6\% and 50.2\%, respectively.
Improving GPU utilization early in execution and achieving consistent scaling across multiple GPUs are key considerations for future work.
\vspace{-0.1cm}
\section{Future Directions}
\niparagraph{Compute-cost analysis.}
Selecting among model families and hyperparameters shifts the Pareto frontier between accuracy and throughput.
Choices like substituting larger ESM variants, using ESMFold or Protenix over alternative predictors, or altering docking seed policies impact both quality and runtime.
Early exploration might prioritize rapid approximations, increasing fidelity later.
Current measurements, however, are input-dependent, complicating comparisons across studies.
We propose standardized metrics such as per-sample latency or energy per accepted candidate at defined quality thresholds.

\niparagraph{Large-scale setups.}
Scaling pipelines introduces complexities as increased sampling or tasks like multimer predictions greatly expand computational demands.
Deploying at scale with multiple GPUs raises challenges in resource placement and scheduling.
Future work should examine packing strategies considering compute and memory jointly and enabling multi-GPU scheduling.
Additional resources like CPU and storage may gain importance.
Investigating interactions between partitioning methods (e.g., MIG, MPS) and length-bucketing strategies to reduce fragmentation is essential.

\niparagraph{Agentic LLMs.}
Expanding candidate sets intensify the bottleneck of human oversight.
Agentic language models (LLMs) can rank candidates, allocate computational resources adaptively, and justify decisions like escalation or early stopping.
The Coscientist~\cite{coscientist:2025}, demonstrating agent-driven scientific discovery, highlights similar orchestration roles in protein-design pipelines.
Integrating LLMs as a new computational component necessitates analyzing their scheduling strategies, resource management policies, and computational behaviors within the existing pipeline infrastructure.

\vspace{-0.1cm}
\vspace{-0.1cm}
\section{Related Work}
To our knowledge, system-level characterizations that profile the full protein design pipeline under realistic co-location are still limited.
At the component level, most works are centered around structure prediction, including characterization~\cite{isswc-alphafold3:2025}, system-level optimization for inference~\cite{nature-colabfold:2022} and training~\cite{megafold:2025}, and an ASIC design~\cite{lightnobel:2025} to support longer sequences.
Homology search has also seen substantial GPU acceleration~\cite{mmseqsgpu}.
Our study complements these efforts by emphasizing end-to-end behavior, resource sharing, and utilization patterns in complete pipelines.

\vspace{-0.1cm}
\section{Conclusion}
\label{sec:conclusion}
We present a reproducible GPU-centric characterization of protein-design pipelines, highlighting computational heterogeneity and GPU underutilization.
Our findings emphasize GPU-aware resource management, cost-aware evaluation metrics, and scalability analyses in multi-GPU setups.
By releasing open-source artifacts, we support future comparable studies and practical pipeline optimizations.

\vspace{-0.1cm}

\bibliographystyle{IEEEtran}
\bibliography{paper}

@article{nextflow,
  title={Nextflow enables reproducible computational workflows},
  author={Di Tommaso, Paolo and Chatzou, Maria and Floden, Evan W and Barja, Pablo Prieto and Palumbo, Emilio and Notredame, Cedric},
  journal={Nature},
  volume={35},
  pages={316--319},
  year={2017}
}

@article{rfdiffusion,
  title={De novo design of protein structure and function with RFdiffusion},
  author={Watson, Joseph L and Juergens, David and Bennett, Nathaniel R and Trippe, Brian L and Yim, Jason and Eisenach, Helen E and Ahern, Woody and Borst, Andrew J and Ragotte, Robert J and Milles, Lukas F and others},
  journal={Nature},
  volume={620},
  pages={1089--1100},
  year={2023}
}

@article{proteinmpnn,
  title={Robust deep learning--based protein sequence design using ProteinMPNN},
  author={Dauparas, Justas and Anishchenko, Ivan and Bennett, Nathaniel and Bai, Hua and Ragotte, Robert J and Milles, Lukas F and Wicky, Basile IM and Courbet, Alexis and de Haas, Rob J and Bethel, Neville and others},
  journal={Science},
  volume={378},
  pages={49--56},
  year={2022}
}

@article{esm-2,
  title={Evolutionary-scale prediction of atomic-level protein structure with a language model},
  author={Lin, Zeming and Akin, Halil and Rao, Roshan and Hie, Brian and Zhu, Zhongkai and Lu, Wenting and Smetanin, Nikita and Verkuil, Robert and Kabeli, Ori and Shmueli, Yaniv and others},
  journal={Science},
  volume={379},
  pages={1123--1130},
  year={2023}
}

@article{protenix,
  title={Protenix-advancing structure prediction through a comprehensive AlphaFold3 reproduction},
  author={{ByteDance AI4Science Team}},
  journal={bioRxiv},
  year={2025},
}

@article{alphafold3,
  title={Accurate structure prediction of biomolecular interactions with AlphaFold 3},
  author={Abramson, Josh and Adler, Jonas and Dunger, Jack and Evans, Richard and Green, Tim and Pritzel, Alexander and Ronneberger, Olaf and Willmore, Lindsay and Ballard, Andrew J and Bambrick, Joshua and others},
  journal={Nature},
  volume={630},
  pages={493--500},
  year={2024}
}

@article{vina,
  title={AutoDock Vina: improving the speed and accuracy of docking with a new scoring function, efficient optimization, and multithreading},
  author={Trott, Oleg and Olson, Arthur J.},
  journal={J. Comput. Chem.},
  volume={31},
  pages={455--461},
  year={2010}
}

@article{vina-gpu,
  title={Vina-GPU 2.1: towards further optimizing docking speed and precision of AutoDock Vina and its derivatives},
  author={Tang, Shidi and Ding, Ji and Zhu, Xiangyu and Wang, Zheng and Zhao, Haitao and Wu, Jiansheng},
  journal={IEEE/ACM Trans. Comput. Biol. Bioinform.},
  year={2024}
}

@inproceedings{diffdock,
  title={DiffDock: Diffusion Steps, Twists, and Turns for Molecular Docking},
  author={Corso, Gabriele and St{"a}rk, Hannes and Jing, Bowen and Barzilay, Regina and Jaakkola, Tommi S.},
  booktitle={ICLR},
  year={2023}
}

@article{nature-primer:2025,
  title={Computational protein design},
  author={Albanese, Katherine I and Barbe, Sophie and Tagami, Shunsuke and Woolfson, Derek N and Schiex, Thomas},
  journal={Nature Reviews Methods Primers},
  volume={5},
  pages={13},
  year={2025}
}

@article{casf-2016,
  title={Comparative assessment of scoring functions: the {CASF-2016} update},
  author={Su, Minyi and Yang, Qifan and Du, Yu and Feng, Guoqin and Liu, Zhihai and Li, Yan and Wang, Renxiao},
  journal={J. Chem. Inf. Model.},
  volume={59},
  pages={895--913},
  year={2018}
}

@article{massivefold,
  title={MassiveFold: unveiling AlphaFold’s hidden potential with optimized and parallelized massive sampling},
  author={Raouraoua, Nessim and Mirabello, Claudio and V{\'e}ry, Thibaut and Blanchet, Christophe and Wallner, Bj{"o}rn and Lensink, Marc F and Brysbaert, Guillaume},
  journal={Nature Computational Science},
  volume={4},
  number={11},
  pages={824--828},
  year={2024}
}

@article{mmseqsgpu,
  title={GPU-accelerated homology search with MMseqs2},
  author={Kallenborn, Felix and Chacon, Alejandro and Hundt, Christian and Sirelkhatim, Hassan and Didi, Kieran and Cha, Sooyoung and Dallago, Christian and Mirdita, Milot and Schmidt, Bertil and Steinegger, Martin},
  journal={Nature Methods},
  pages={1--4},
  year={2025},
}

@article{nature-opportunities:2024,
  title={Opportunities and challenges in design and optimization of protein function},
  author={Listov, Dina and Goverde, Casper A and Correia, Bruno E and Fleishman, Sarel Jacob},
  journal={Nature},
  volume={25},
  pages={639--653},
  year={2024}
}

@article{coscientist:2025,
  title={Towards an AI co-scientist},
  author={Gottweis, Juraj and Weng, Wei-Hung and Daryin, Alexander and Tu, Tao and Palepu, Anil and Sirkovic, Petar and Myaskovsky, Artiom and Weissenberger, Felix and Rong, Keran and Tanno, Ryutaro and others},
  journal={arXiv},
  year={2025},
}

@inproceedings{lightnobel:2025,
  title={LightNobel: Improving Sequence Length Limitation in Protein Structure Prediction Model via Adaptive Activation Quantization},
  author={Han, Seunghee and Choi, Soongyu and Kim, Joo-Young},
  booktitle={ISCA},
  pages={1940--1955},
  year={2025}
}

@inproceedings{isswc-alphafold3:2025,
  title={AlphaFold3 Workload Characterization: A Comprehensive Analysis of Bottlenecks and Performance Scaling},
  author={Kim, Jinpyo and Kwon, Mingi and Zhao, Jishen},
  booktitle={IISWC},
  year={2025}
}

@article{megafold:2025,
  title={MegaFold: System-Level Optimizations for Accelerating Protein Structure Prediction Models},
  author={Hoa La and Ahan Gupta and Alex Morehead and Jianlin Cheng and Minjia Zhang},
  journal={arXiv},
  year={2025},
}

@article{nature-colabfold:2022,
  title={ColabFold: making protein folding accessible to all},
  author={Mirdita, Milot and Sch{"u}tze, Konstantin and Moriwaki, Yoshitaka and Heo, Lim and Ovchinnikov, Sergey and Steinegger, Martin},
  journal={Nature methods},
  volume={19},
  pages={679--682},
  year={2022}
}

\vfill

\end{document}